\documentclass[preprint,journal]{IEEEtran}

\usepackage{graphicx}
\usepackage[numbers,sort&compress]{natbib}
\usepackage[dvipsnames]{xcolor}
\ifCLASSINFOpdf
\else
\fi
\usepackage{amsmath,amssymb}
\begin{document}
© 20XX IEEE.  Personal use of this material is permitted.  Permission from IEEE must be obtained for all other uses, in any current or future media, including reprinting/republishing this material for advertising or promotional purposes, creating new collective works, for resale or redistribution to servers or lists, or reuse of any copyrighted component of this work in other works.

%
\title{Dynamic Processing Neural Network Architecture For Hearing Loss Compensation}
%
%
%

\author{Szymon Drgas,
        Lars Bramsl\o w,
        Archontis Politis,
        Gaurav Naithani,
        Tuomas Virtanen
\thanks{S. Drgas is with Institute of Automatic Control and Robotics, Poznan University of Technology, Poznan, Poland, e-mail: szymon.drgas@put.poznan.pl}
\thanks{L. Bramsl\o w is with Eriksholm Research Center, e-mail: labw@eriksholm.com}    
\thanks{T. Virtanen, A. Politis, and G. Naithani are with Faculty of Information Technology and Communication Sciences, Tampere University, Tampere, Finland, e-mail fistname.lastname@tuni.fi}
\thanks{Manuscript received April 19, 2005; revised August 26, 2015.}}

\maketitle
\thispagestyle{empty}
\pagestyle{empty}
\begin{abstract}
This paper proposes neural networks for compensating sensorineural hearing loss. The aim of the hearing loss compensation task is to transform a speech signal to increase speech intelligibility after further processing by a person with a hearing impairment, which is modeled by a hearing loss model.
We propose an interpretable model called dynamic processing network, which has a structure similar to band-wise dynamic compressor. The network is differentiable, and therefore allows to learn its parameters to maximize speech intelligibility. More generic models based on convolutional layers were tested as well.
The performance of the tested architectures was assessed using spectro-temporal objective index (STOI) with hearing-threshold noise and hearing aid speech intelligibility (HASPI) metrics. 
The dynamic processing network gave a significant improvement of STOI and HASPI in comparison to popular compressive gain prescription rule Camfit. A large enough convolutional network could outperform the interpretable model with the cost of larger computational load. Finally, a combination of the dynamic processing network with convolutional neural network gave the best results in terms of STOI and HASPI.
\end{abstract}


%
\IEEEpeerreviewmaketitle

\section{Introduction}
Recent reports \cite{chadha2021world,10665-339913} indicate that demographic and population trends reflect the high, and rising, global
prevalence of hearing loss across the life course. It is estimated that by 2050, about 2.5 billion (1 in every 4) people will experience hearing loss, with nearly 700 million (1 in every 14) living with moderate or higher levels of hearing loss in the better hearing ear. Hearing loss can negatively affect language development, psychosocial well-being, quality of life, educational attainment and economic independence at various stages of life \cite{nordvik2018generic}.
According to \cite{lesica2021harnessing}, one of the top challenges for artificial intelligence related to hearing is development of effective machine learning-based signal processing solutions addressing hearing loss, with over 100 million people currently in need of one.
    
The negative aspects of hearing loss originate to a large extent from reduced speech intelligibility, especially in adverse acoustic environments. This is due to hearing loss inducing distortions of neural activity patterns \cite{lesica2018hearing}. Such distorted neural patterns elicited by different sounds are less unique and less robust to background noise.

A core strategy in hearing aids is some form of amplification, since audibility is a necessary – but not a fully sufficient – requirement for better speech intelligibility. In the simplest form, linear amplification applying the same amount of gain at all input levels is appropriate for all hearing
impairments not being associated with dynamic range reduction, such as conduction losses or pure inner hair cell loss resulting in
rare cases of sensorineural impairment without recruitment.
However,
most hearing impaired patients have sensorineural impairments with reduced dynamic range, predominantly due to outer hair cell loss, which requires non-linear amplification to amplify soft sounds with higher gains compared to loud sounds, i.e. audio compression. 
There are different types of non-linear amplification schemes \cite{kates2005principles} which can be classified according to different criteria (slow/fast, input-/output controlled, single-/multi-channel compression).
Modern non-linear amplification schemes are typically implemented as multi-channel wide dynamic range compression (WDRC) systems to allow frequency-specific loudness restoration.
\cite{kollmeier2018functionality}. Although multi-channel compression in hearing aids can improve quality of life indices \cite{chisolm2007systematic}, \cite{larson2000efficacy}, this approach enhances the perception of weak sounds, but, unfortunately, is not sufficient to restore the perception of speech in noisy environments \cite{humes1999comparison}. This is because current hearing aids are designed to restore neural activity to its original level. Hearing loss, however, does not simply weaken neural activity, it significantly distorts the neural patterns.

Deep neural networks have potential to compensate the distortion of the neural patterns caused by hearing loss. One possibility is to train a neural network to compensate hearing loss using speech signals and corresponding recordings of neural responses gathered by means of intrusive techniques like a probe with hundreds of microelectrodes \cite{shobe2015brain}. However, such responses in large amounts may be difficult to record, especially as the distortions are listener-dependent.

Another possibility is to train deep learning models, that transform a speech signal, to make it as intelligible as possible, for a given  (possibly hearing-impaired) listener. The parameters of the hearing loss compensation model can be optimized with a loss function that assesses  intelligibility of the processed signal. Such a loss function can be constructed using two main components: (1) a hearing loss model that discards information from the signal that is not available to a given hearing impaired listener (for example inaudible cues) (e.g. \cite{jepsen2008computational,carney2017fluctuation,zilany2006modeling}), (2) speech intelligibility prediction, which reflects how many words the listener would understand from the signal.

In \cite{tu2021DHASP} a  trainable finite impulse response (FIR) filter was used as a differentiable hearing aid processor for clean signals. It was trained using a loss function that compared signal envelopes processed with models of normal and impaired hearing. These envelopes were compared using a cepstral correlation measure \cite{kates2014hearing} and an energy control function. The performance of the methods was measured using hearing aid speech intelligibility (HASPI) \cite{kates2014hearing}. The trainable FIR outperformed a hearing loss compensation method that applied fixed gains in frequency bands set according to NAL-R - the revised prescriptive procedure to fit a hearing aid from the National Acoustics Laboratory Australia (NAL) \cite{byrne1986national}. This formula gives frequency-specific gains which are determined so that frequency components of a speech signal are presented with approximately equal loudness without exceeding comfort level. This work was further extended for noisy conditions in \cite{tu2021optimising}, using magnitude spectrogram differences instead of envelope differences as the loss function. 
It was concluded that the data-driven optimized fitting can outperform the prescribed ones especially for mild and moderate hearing loss. It was also concluded that noise-dependent optimization is promising.

In general, deep neural networks offer the ability to optimize complex nonlinear signal processing according to objective functions that reflect speech intelligibility \cite{desloge2012auditory}; a task that is uniquely suited to processing for hearing aids, considering the non-linear nature of hearing loss.
In the Clarity Challenge 2021 \cite{graetzer2021clarity} where binaural signals were simulated in noisy reverberant conditions for various listeners, two submissions optimized the hearing-loss compensation according to the speech intelligibility metric. In \cite{tu2021atwostage} an amplification module was trained using the STOI objective to compensate the hearing loss model. The amplification module was either a trainable FIR or a Conv-TASNet DNN architecture \cite{luo2019conv}. 
Similarly to \cite{tu2021atwostage}, a Conv-TASnet network was used as a separator of the target speaker from noise and interference in \cite{zmolikova2021BUT}. However, after training the separator module, the authors of \cite{zmolikova2021BUT} inject an auxiliary network of three fully connected layers with leaky ReLU activations, between the encoder and decoder parts of the separator. The purpose of the auxiliary network is to personalize the model to a certain user's hearing loss, by taking their audiogram as input and producing time-variant gains to be applied in the separator's encoded features, before passing them to the decoder. In \cite{drakopoulos2022differentiable} a differentiable optimization framework for DNN-based hearing aid strategies was proposed. In this case the hearing loss model is a neural network trained to simulate model. This modeling is focused on cochlear synaptopathy (damage to auditory-nerve synapses) and hence it does not address HL greater than 30 dB HL, which has limited significance in a hearing aid context. Another attempt to train a neural network for hearing loss compensation is in \cite{cheng2023speech}. The convolutional encoder-decoder network was trained to maximize hearing-aid speech quality index (HASQI) computed by a appropriately trained network.

According to the literature mentioned above, the recent differentiable hearing aid architectures are mainly linear amplification modules, or fairly large neural networks (e.g. Conv-TASNet). Those studies (\cite{tu2021DHASP,tu2021atwostage,luo2019conv,zmolikova2021BUT}) seem more targeted towards showing the benefits of data-driven trainable models and differentiable signal processing in general, in comparison to classical hearing aid processing, rather than studying how to strongly couple the differentiable architecture to the hearing loss problem, or how architectural choices or the parameters of those models affect performance in detail. Differentiable architectures motivated by the structure of the hearing loss model can be less complex (in terms of the number of arithmetic operations) than generic DNNs. Furthermore, as hearing loss models are systems consisting of many components (e.g. spectral smearing, loudness recruitment, dead regions), it is important to analyze how the individual components of the hearing loss model can be compensated. To our best knowledge, data-driven compensation against individual components of the hearing loss model have not been studied before. Additionally, the network parameters are optimized using losses of similarity between signal envelopes, but the audibility of these signals is not taken into account.

In this work we address the problem of designing learnable models that process the speech signal to compensate the hearing loss. We consider two trainable systems as hearing loss compensation processors: a) a hearing loss-informed differentiable multi-band dynamic processor and b) general-purpose convolutional neural networks. We systematically evaluate the influence of architectural choices on objective intelligibility metrics. These architectural choices cover the extensions of the dynamic processing network in which more aspects of processing are tuned (additional filters for processing of envelope and parameters of filterbanks) and hyperparameters of CNNs (number of layers and the size of kernels of CNN layers).  We also test the ability of the proposed models to compensate the individual components (i.e. spectral smearing and loudness recruitment) of the hearing loss model. 
The proposed architectures are trained using a modified STOI loss function, in which inaudible sounds do not affect the predicted speech intelligibility. This prevents overestimating STOIs by increasing positive correlations of compared signals below the hearing threshold.

The structure of the paper is as follows. In Section \ref{sec:hl_comp} the general description of training of hearing loss compensation is described. Next, in Section \ref{sec:hlmodel}, our implementation of the differentiable Moore-Stone-Baer-Glasberg (MSBG) model \cite{nejime1997simulation} is presented. This is followed by the description of architectures of trainable hearing loss compensation networks in Section \ref{sec:architectures}. The experimental evaluation is presented in Section \ref{sec:experiments}. Finally, conclusions are drawn in Section \ref{sec:conclusions}. 

\begin{figure*}[th]
    \centering
    \includegraphics[width=12cm]{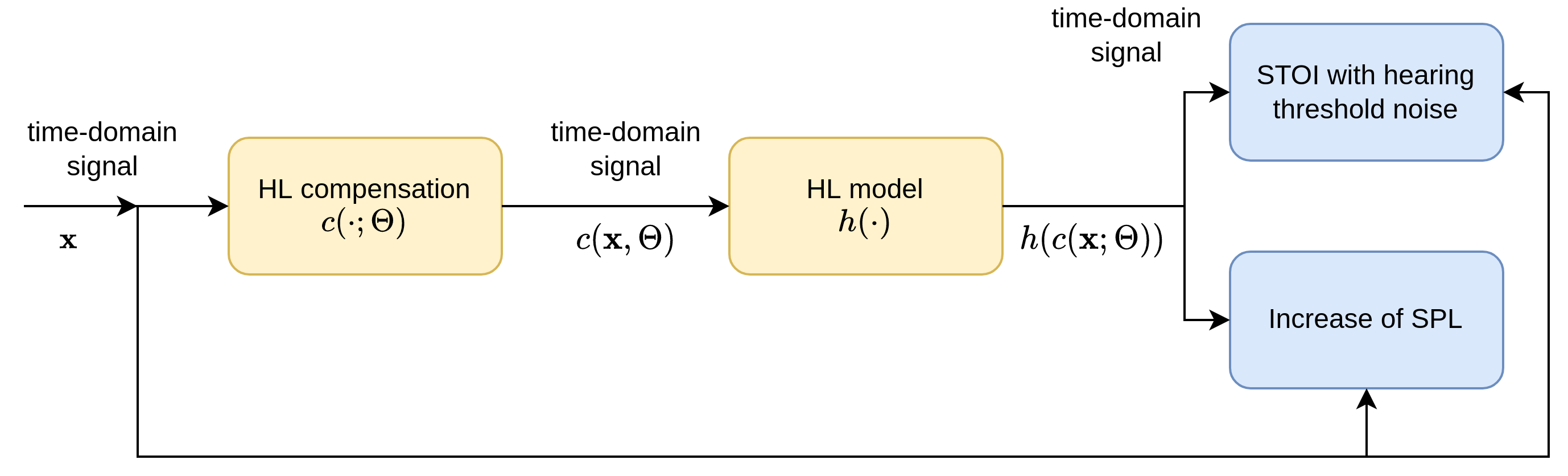}
    \caption{Overview of the training of hearing loss compensation system}
    \label{fig:overview}
\end{figure*}

\section{Hearing loss compensation}
\label{sec:hl_comp}
The overview of the proposed hearing loss compensation system is presented in Figure \ref{fig:overview}. It corresponds to a monaural hearing scenario in which audio signal is processed by the hearing aid and then fed to the ear of the user, who has hearing loss. Therefore, the time-domain input signal is first processed by the hearing loss compensation network. Next, the processed signal is degraded using the hearing loss model, which is a differentiable version of the model presented in \cite{nejime1997simulation} (see Section \ref{sec:hlmodel} for details). 

The transformation of the input signal by the compensation network and the hearing loss model can be written as
\begin{equation}
    {\bf y} = h(c({\bf x};\Theta))=(h\circ c)({\bf x};\Theta)\;
\end{equation}
where ${\bf x}$ is a vector with time-domain samples of the input signal, $h$ is the function simulating hearing loss, and $c$ is the hearing loss compensation function (neural network) with a set of parameters $\Theta$. 

\subsection{Loss function}
STOI is an objective metric that is related to speech intelligibility. It is based on the correlation of envelopes of the compared signals in one-third octave bands, and therefore, does not depend on the level of the signals. Without any additional criteria, maximizing STOI will lead to increased level of the output signal. Therefore, we will use a loss function that consists of STOI and an additional term that prevents the output of the hearing loss model to have higher level than the level of the clean speech signal. Thus, the parameters of the compensation network $c$ are optimized by minimizing the loss function
\begin{equation}
    \label{eq:loss}
    \mathcal{L}=-\alpha\textrm{STOI}({\bf x}, {\bf y}) + (1-\alpha)\max(0, L({\bf y})-L(\bf x))
\end{equation}
with respect to parameters $\Theta$, where $\textrm{STOI}(\cdot,\cdot)$ is a spectro-temporal objective intelligibility function (see Section \ref{sec:metrics} for more details), $L(\cdot)$ is the level of a given signal (e.g. for $N$-component vector ${\bf x}$ the level is computed as $L({\bf x})=10\log_{10}(\|{\bf x}\|^2)/N$), and $\alpha\in[0,1]$ is the weight for balancing the influence of both loss terms. 

\subsection{STOI with hearing threshold noise}
\label{sec:thrnoise}
The hearing loss model can also decrease signal's level in frequency bands, such that it is below the hearing threshold. Since STOI is level-independent, a signal that is below the hearing loss threshold can be still correlated with the reference signal and contribute positively to the predicted speech intelligibility. In order to overcome this effect, we propose to add hearing-threshold noise to the processed signal (after the hearing loss model) before computation of STOI. The similar solution is done for example in HASPI metric~\cite{kates2005principles}.

The hearing-threshold noise is obtained by filtering Gaussian noise with FIR filter which frequency characteristics is based on MAP (minimum audible pressure) curve taken from \cite{killion1978revised} (using \verb|firwin2| function from scipy python package). Next, the overall level of the whole noise is scaled in such way, that the level at the output of third octave filter with center frequency 1000 Hz is 15 dB higher that the level of MAP. These additional 15 dB come from the observation, that the difference that is needed to mask a tone with a band-pass noise is about 15 dB.
    
\section{Differentiable hearing loss model}
\label{sec:hlmodel}
The hearing loss model used in this study is based on the model presented in \cite{nejime1997simulation}\footnote{https://github.com/claritychallenge}. There are two main parts of the model, namely spectral smearing, which blurs the magnitude spectrum, and loudness recruitment, which performs band-wise dynamic expansion \cite{cai2009encoding}.

Before the above stages, the input time-domain signal is filtered using a 1000-tap FIR filter as
\begin{equation}
    {\bf x}_{\rm coch}={\bf x}*{\bf f}_{\rm to\_coch}\;,
\end{equation}
where $*$ denotes convolution of discrete signals, whose samples are components of the operand vector and ${\bf f}_{\rm to\_coch}$ is a vector comprising the impulse response of the filter that represents characteristics of outer and middle ear, based on the model of \cite{shaw1974transformation} (see Appendix for more information). 

\subsection{Spectral smearing}
One of the consequences of the damage of the cochlea is spectral smearing. It is connected to the broadening of the auditory filters. Consequently, the auditory simultaneous masking is stronger, as the broader filters can transmit more noise and degrade the signal-to-noise ratio. Additionally, broadening of the auditory filters causes a decrease in spectral resolution and therefore the ability to accurately recognize speech and music sounds by a listener. 

In the hearing loss model, magnitude spectra are smeared in the frequency dimension. The spectral smearing simulation begins with the calculation of the STFT as
\begin{equation}
    {\bf X}={\rm STFT}({\bf x}_{\rm coch})\;,
\end{equation}
where ${\bf X}\in\mathbb{C}^{B\times T_{\rm smear}}$ and $B$ is the number of bands (the number of FFT bands divided by two plus one) and $T_{\rm smear}$ is the number of frames. The STFT is calculated with hop length of 64 samples ($\approx 1.5$ ms) and Hann windowing of length 256 ($\approx 6 $ ms). The windowed signal is transformed using 512-point fast Fourier transform (FFT). Next, magnitude ${\bf M}$ and phasor ${\bf P}$ matrices are calculated from ${\bf X}$ as
\begin{equation}
    {\bf M}=|{\bf X}|\;,
\end{equation}
where $|\cdot|$ denotes entry-wise absolute value and
\begin{equation}
    {\bf P}={\bf X}\oslash({\bf M}+\epsilon)\;,
\end{equation}
where $\oslash$ is entry-wise division and $\epsilon$ is a very small value to prevent division by zero. The magnitude spectrum is smeared using a smearing matrix ${\bf S}$ as
\begin{equation}
    \label{eq:smearing}
    {\bf M}_{\rm smeared}=\sqrt{{\bf SM}^2}\;,
\end{equation}
where $\sqrt{\cdot}$ and $\cdot^2$ are done entry-wise.
Matrix ${\bf S}$ is constructed in such a way that the operation in (\ref{eq:smearing}) broadens the auditory filters  for frequencies lower than center frequency by a factor of 4, and by a factor of 2 for frequencies higher than center (see Appendix for more details).
Finally, the time-domain smeared-spectrum signal is calculated as 
\begin{equation}
    {\bf x}_{\rm smeared}={\rm iSTFT}({\bf M}_{\rm smeared}\odot {\bf P})\;,
\end{equation}
where $\odot$ denotes entry-wise multiplication and $\textrm{iSTFT}(\cdot)$ is the inverse STFT.

\subsection{Loudness recruitment}
\label{sec:lr}
Weak sounds do not drive the basilar membrane strongly enough to elicit auditory nerve activity. An active amplification by outer hair cells (OHC) reinforce the passive movement of the basilar membrane which allows perception of sounds with low intensity. The amplification provided by the OHCs decreases as the level of sound increases. Therefore, the active mechanism provides compression of the sound. If the cochlea is damaged, an abnormal increment of loudness with the sound pressure level is perceived, as for low intensities sound is inaudible while for higher intensities it has normal loudness. This phenomenon is called loudness recruitment.

The loudness recruitment is modeled as dynamic expansion performed in frequency bands. In this work, the signal is filtered into bands using a gammatone filterbank. The signal envelope in each frequency band is used to control the band-dependent gain that is used to multiply the sub-band signals. Finally, the dynamically processed sub-band signals are mixed.

As a first step, the filtering by the gammatone filterbank is realized as  
\begin{equation}
    \label{eq:lrin}
    {\bf x}_c={\bf x}_{\rm smeared}*{\bf g}_c\;,
\end{equation}
for $c=1,\ldots,C_{\rm recruit}$, where ${\bf g}_1,\ldots,{\bf g}_{C_{\rm recruit}}$ are the impulse responses of gammatone filters (first 2000 samples of each filter) and $C_{\rm recruit}$ is the number of frequency bands used in the loudness recruitment model. This step corresponds to the spectral analysis done by the basilar membrane.
In the next step, envelope of each frequency band is computed by fully-rectifying the band signal (taking the absolute value of each sample resulting to $|{\bf x}_c|$) and low-pass filtering the result. For computational efficiency, envelopes are downsampled and any further processing is performed with their downsampled representations. Downsampling of the fully-rectified signal in each band $c$ is done with an 1D convolutional layer and a stride equal to 50 samples, with the kernel set to an antialiasing filter ${\bf f}_{\rm aa}$  which is a low-pass FIR filter of order 1660 with cut-off frequency of $\frac{\pi}{50}\frac{{\rm rad}}{\rm sample}$, designed with the window method using a Hann window. The downsampled envelope ${\bf e}_c$ signal is computed by the convolution of $|{\bf x}_c|$ with ${\bf f}_{aa}$ followed by decimation with the factor of 50. 
Following downsampling, each band ${\bf e}_c$ is individually filtered by a channel-specific low-pass filter ${\bf l}_c$.
Filters ${\bf l}_1,\ldots,{\bf l}_{C_{\rm recruit}}$ have a cutoff frequency $f_c$ equal to the width $(3/4)\min(100, f_c)$ of the corresponding gammatone filter (see Appendix for more information). 
After the low-pass filtering, the signals are upsampled back to the original sampling rate by creating a sequence from ${\bf e}_c$ comprising the samples of the downsampled envelope, each of them separated by $L_1$ zeros, and filtering the sequence by ${\bf f}_{aa}$.
The resulting envelope is clipped, i.e. its values above $e_{\rm max}$ are changed to $e_{\max}$, where $e_{\rm max}$ is the envelope value that corresponds to 105 dB SPL.
Next, a gain signal is computed as
\begin{equation}
    g_c(n)=\left(\frac{\tilde{e}_c(n)}{e_{\rm max}}\right)^{\gamma_c-1}\;,
\end{equation}
where $e_c(n)$ is the $n$'th component of ${\bf e}_c$ and
\begin{equation}
    \gamma_c = \frac{\eta}{\eta-h_c}\;,
\end{equation}
where variable $\eta=105$ is the catch-up parameter, and $h_c$ is the audiogram value for frequency channel $c$.
The computed gain $g_c(n)$ is used to multiply the samples of the signal $x_c(n)$, i.e.
\begin{equation}
    y_c(n)=x_c(n)g_c(n),
\end{equation}
in each band $c$. The gain usually attenuates low-level sounds, what reflects the dysfunction of the OHCs. Finally, the processed frequency bands are mixed and multiplied by the recombination gain $r=10^{\frac{4.24}{20}}$ as
\begin{equation}
    y(n)=r\sum_{c=1}^{C_{\rm recruit}} y_c(n)\;.
\end{equation}
After the loudness recruitment the inverse filter ${\bf f}^{-1}_{\rm to\_coch}$ is applied (see Appendix for more details). This is followed by low-pass filter with cutoff frequency of 18000 Hz to suppress non-linear high-frequency artifacts.

\section{Architectures for hearing loss compensation}
\label{sec:architectures}
In this section, we propose neural architectures for the hearing loss compensation (see $c(\cdot;\Theta)$ in Figure \ref{fig:overview}). There are several desired conditions that the compensation network should fulfill. First, in practical applications the hearing-loss compensation network should be causal, and the latency introduced by this network should be smaller than 10 ms \cite{stone2008tolerable}. In this work we address these requirements by using causal convolutional layers in our models and short-time Fourier transform (STFT) windows of 5 ms.  Additionally, the complexity of the compensation network should be limited, as it is related to processing capabilities of the processor and power consumption in a potential hearing aid. In the proposed dynamic processing network, the number of arithmetic operations is relatively low in comparison to generic DNN models. For example, the proposed dynamic processing network, in its basic form performs 12874 arithmetic operations per spectrogram frame, while one level of 2D convolutional layer (used in the experiments - 48 filters with $7\times 3$ kernel, accepting feature map with 48 features for each time-frequency unit), performs over 12 million arithmetic operations per spectrogram frame.
\subsection{Dynamic processing network}
\label{sec:dp}
\begin{figure}
    \centering
    \includegraphics[width=5cm]{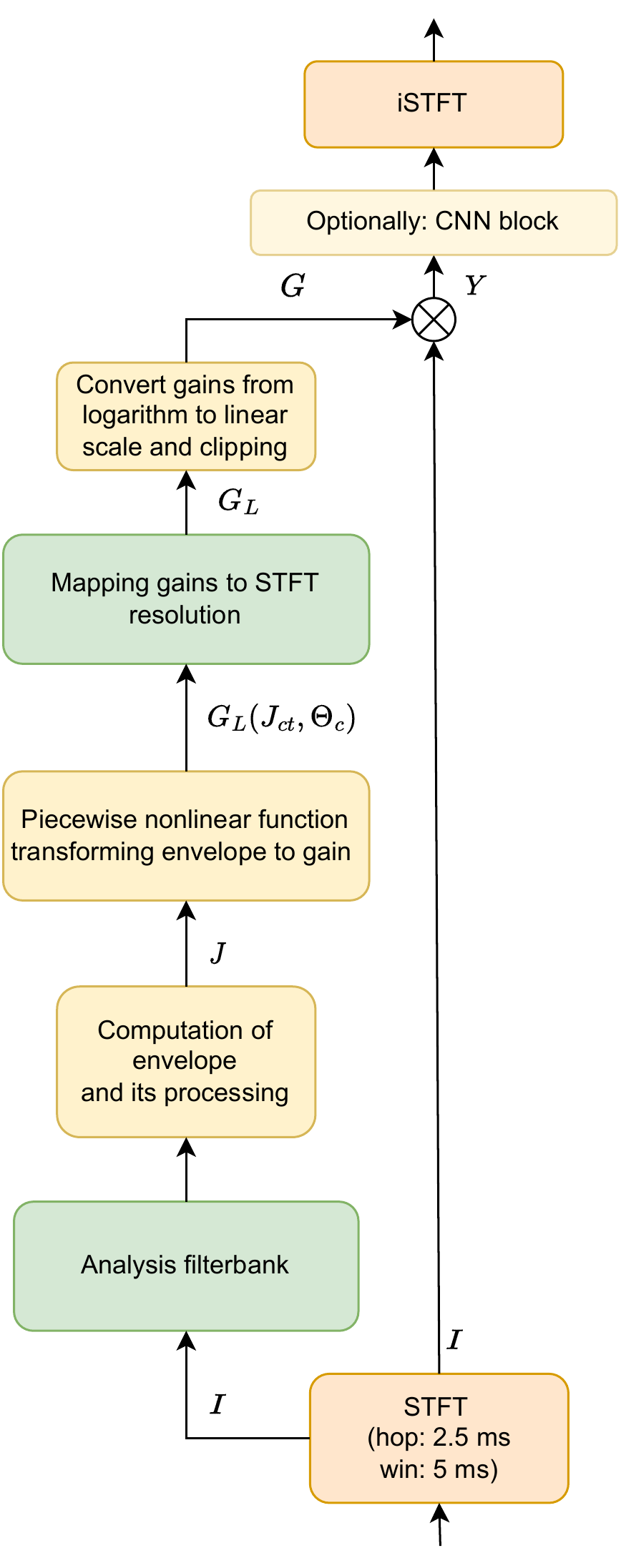}
    \caption{The structure of the proposed dynamic processing network}
    \label{fig:dpn}
\end{figure}

In this paper we propose a dynamic processing network, which is a differentiable computational model that can perform dynamic processing of a signal in frequency bands. In dynamic processing, a signal is amplified or attenuated using a gain variable in time, which is dependent on the RMS calculated from the signal locally in time. The parameters of the dynamic processing network can be set according to  gain tables which can be determined for example with Camfit procedure \cite{moore2010development}, in which the configuration of the multi-band compressor is obtained according to a model of loudness perception. Next, these parameters can be further fine-tuned,  to maximize the predicted intelligibility of the processed signal based on loss~(\ref{eq:loss}). Thus, in the basic form of the dynamic processing network, its parameters are initialized according to Camfit using the procedure described in \ref{sec:caminit}, and next they are fine-tuned. 

The general architecture of the dynamic processing network is shown in Figure \ref{fig:dpn}. First, the STFT of the input to the dynamic processing network is computed. Next, the processing is split in two branches. The left branch in Figure \ref{fig:dpn} is calculating gains based on envelopes of signals in frequency bands, which are then used to band-wise multiply signals from the right branch. The gains and the envelopes are calculated in frequency bands obtained using the analysis filterbank and later they are transformed-back to gains for the STFT bands using the synthesis filterbank. Finally the inverse STFT is used to get the time domain signal. The analysis and synthesis filterbanks are used to reduce the number of frequency bands, and therefore the number of parameters, as in each frequency band a function transforming envelopes to gains is learned. 

A gain is a piece-wise linear function of an envelope, and it depends on parameters that can be tuned using a learning procedure. In the basic form, only parameters of the piece-wise linear function are fine-tuned. We also consider extensions in which parameters of the envelope processing filters (see Section \ref{sec:envproc} and the filters of the analysis and the synthesis filterbanks from Figure \ref{fig:dpn} are fine-tuned as well. 

The STFT block in Figure \ref{fig:dpn} computes tensor $I\in\mathbb{R}^{B\times T_{\rm comp}\times 2}$ of the input signal
 ${\bf x}$, where $B=N_{\rm FFT}/2 + 1$, $T_{\rm comp}$ is the number of frames, and the last dimension corresponds to the real and imaginary parts of the spectrogram.

For gain computation, the STFT tensor is decimated to $C$ frequency bands. The RMS envelope in frequency band $c$ for time frame $t$ is calculated as
\begin{equation}
    \label{eq:fbanal}
    J_{ct}=\sqrt{{\frac{2}{N_{\rm FFT}}}\sum_{b=1}^B F_{cb}(I_{bt1}^2 + I_{bt2}^2)}\;,
\end{equation}
where $F_{cb}$ consists of the response of $c$'th analysis filter for each STFT frequency $b$. A response $F_{cb}$ is equal to one for STFT frequencies $b$ between cutoff frequencies of band $c$ and zero otherwise. The cutoff frequencies for bands $c=1,\ldots,C$ are spaced logarithmically. 
\begin{figure}
    \centering
    \includegraphics[width=8cm]{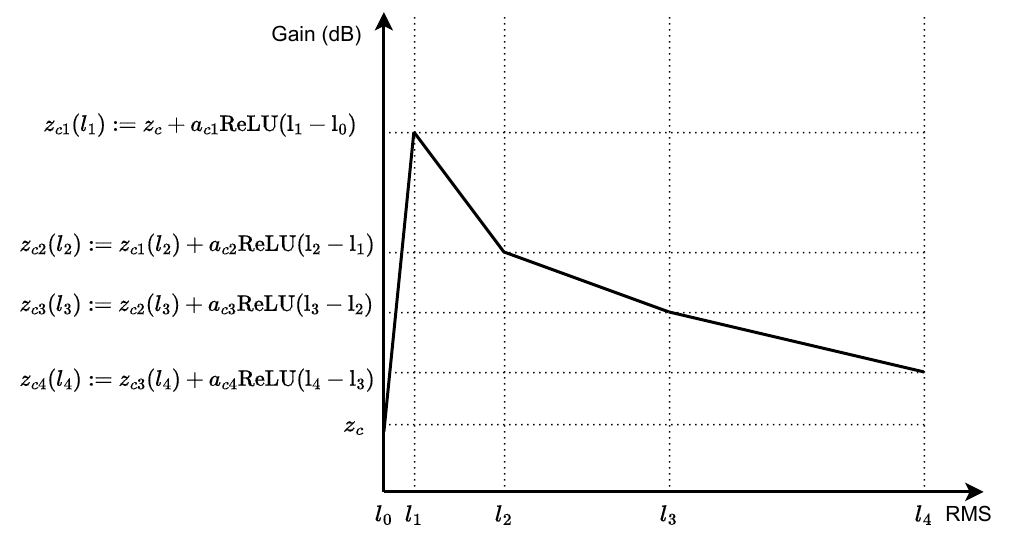}
    \caption{Schematic illustration of piece-wise linear function used in the dynamic processing network}
    \label{fig:piecewise}
\end{figure}
The envelopes can be optionally processed by smoothing filters as presented in Section IV-A2.

Gain values in each channel $c=1,\ldots,C$ are calculated based on a piece-wise linear function as
\begin{equation}
    \label{eq:piecewise}
    G_L(J_{ct}, \Theta_c) = \sum_{i=1}^{H}a_{ci}{\rm ReLU}(J_{ct}-l_{i-1}) + z_c\;,
\end{equation}
where $\textrm{ReLU}(x)=\max(x,0)$. Equation (\ref{eq:piecewise}) represents the dependence between RMS (linear scale) of a signal in band and its amplification (dB scale). The slopes of segments of the piece-wise linear function depend on learnable parameters $\Theta_c = (a_{c1},\ldots,a_{cH},z_c)$,
and $l_0,\ldots,l_H$ are fixed RMS values that are boundaries between subsequent pieces of (\ref{eq:piecewise}). In our experiments $l_0,\ldots,l_H$ are RMS values which correspond to levels from -10 to 111 dB SPL with 1 dB step.
The piece-wise linear function from Equation (\ref{eq:piecewise}) is shown in Figure \ref{fig:piecewise}. 

The gain (in dB) for STFT band $b$ and time frame $t$ is calculated by setting it to  ${G_L}_{J_{ct}}$ for frequency band $c$ which has $b$ in its pass-band:
\begin{equation}
    \label{eq:fbsynth}
    {G_L}_{bt}=\sum_{c=1}^C F_{cb}G_L(J_{ct};\Theta_c)\;,
\end{equation}
and transformed to linear scale as
\begin{equation}
    G_{bt}=10^{{G_L}_{bt}/20}\;.
\end{equation}
Next, the gains $G_{bt}$ are clipped, to limit the maximal gain to 10000 (80 dB). Finally, the input signal in the STFT domain is multiplied by the corresponding gain $G_{bt}$ as
\begin{equation}
    \label{eq:outputstft}
    Y_{btk} = I_{btk}G_{bt}\;,
\end{equation}    
for $b=1,\ldots, B$, $t=1,\ldots,T_{\rm comp}$ and $k=1,2$, where $k=1$ refers to real part while $k=2$ to the imaginary part of spectrograms $Y$ and $I$. Tensor $Y$ can be optionally processed by CNN block (marked in Figure \ref{fig:cnn}). 
\subsubsection{Initialization of dynamic processing network with Camfit gain tables}
\label{sec:caminit}
\begin{figure}
    \centering
    \includegraphics[width=9cm]{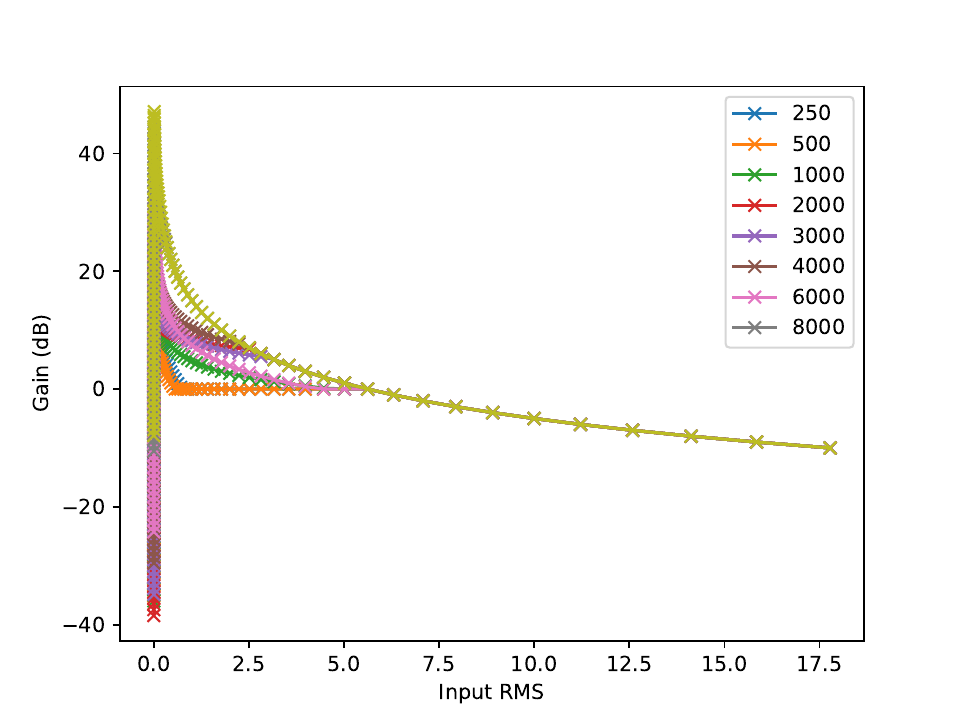}
    \caption{Gain tables obtained using Camfit procedure (1 RMS = 85 dB SPL)}
    \label{fig:Camfit}
\end{figure}

The parameters of the dynamic processing network presented in the previous section can also be obtained from the Camfit procedure \cite{moore2010development}. As the dynamic processing network is differentiable, the initialized parameters can be additionally fine-tuned according to the loss function (\ref{eq:loss}). The Camfit procedure allows to determine the characteristics of dynamic compression in a hearing aid according to an audiogram only. The parameters of the compressor are derived to recover the perception of loudness for a hearing-aided hearing impaired listener. The result of the Camfit is often provided in the form of gain tables $L_{cj}$, which indicate the gain for each frequency band $c$ and input RMS with index $j$. Example of curves plotted according to gain tables are shown in Figure \ref{fig:Camfit}.  This subsection shows how to transform the gain tables to the parameters ($\Theta_1,\ldots,\Theta_C$) of the proposed dynamic processing network. 
In order to express gain tables as a combination of shifted $\textrm{ReLU}$ functions (see Equation (\ref{eq:piecewise})) $z_c=L_{c0}$ and the parameters $a_{ci}$  can be calculated as 
\begin{equation}
    [a_{c1}\;\ldots\; a_{cH}]^{\rm T}={\bf A}^{-1}\left[\begin{array}{c}
         L_{c1}-z_c  \\
         \vdots \\
         L_{cH}-z_c
    \end{array}\right]
\end{equation}
where 
\begin{equation}
    A_{ij}={\rm ReLU}(l_i - l_{j-1})\;,
\end{equation}
for $i=1,\ldots,H$ and $j=1,\ldots,H$.
The input values in $J_{ct}$ are scaled, such that they represent RMS corresponding to SPL.

\subsubsection{Envelope processing}
\label{sec:envproc}
We also study processing of envelopes that control the gains in frequency bands. We study two types of processing: FIR filters which coefficients are learnable, and 1-st order IIR filters with fixed parameters. The filtering can result in removing of unwanted patterns/frequencies from envelopes and therefore increase the intelligibility of the signal at the output of the dynamic processing network. 

The learnable FIR filter is expressed as
\begin{equation}
    \bar{J}_{ct} = \sum_{\tau=0}^{\mathcal{T}}w_c(\tau)J_{c(t-\tau)}\;,
\end{equation}
for $c=1,\ldots,C$ and $t=1,\ldots,T$, where $w_c(1),\ldots,w_c(\mathcal{T})$ are the coefficients of the envelope filter for frequency band $c$, and $\mathcal{T}$ is the length of the filters.

The studied IIR filter is typical for hearing aids, i.e. filter with 10 ms of attack and 100 ms of release time. This was implemented using a recurrent layer, that computes 1-st order IIR filter, with coefficient $\tilde{w}=\exp(-T/\tau)$ where $\tau$ is set  to $\tau_{\rm attack} = 10$ ms when the current input is greater than the output from the previous step and $\tau_{\rm release}=100$ ms otherwise. The envelope smoothing can be described with the following steps for input $j_{ct}$ of channel $c$ and RNN (recurrent neural network) cell with state $s_{c,t}$:
\begin{equation}
    \Delta_{ct} =  s_{c, t-1}-j_{ct}\;,
\end{equation}
\begin{equation}
    \bar{w}_{ct}=\left\{\begin{array}{ccc}
        \tilde{w}_{\rm attack} & & \Delta_{ct} \geq 0 \\
        \tilde{w}_{\rm release} & & \textrm{otherwise} 
    \end{array}\right.\;,
\end{equation}
\begin{equation}
    s_{ct}=s_{c,t-1} + (j_{ct}-s_{c,t-1})\bar{w}_{ct}\;.
\end{equation}
State $s_{ct}$ is the smoothed output corresponding to the current input frame $j_{ct}$.

\subsubsection{Audiogram-controlled compensation}
\label{sec:audiogramdepend}
In the basic form of the dynamic processing network its parameters $\Theta_1,\ldots,\Theta_C$ are trained for each listener individually. This can be time- and hardware- consuming. We also consider a listener-independent variant of the compensation network that is trained jointly for all the listeners, but which takes an audiogram of a listener as an additional input, to produce listener-specific compensation. In the listener-independent variant of the dynamic processing network, the coefficients $\Theta_c=(a_{c1},\ldots,a_{cH},z_c)$ are changed to be a function  
\begin{equation}
    \Theta_c(h_c) = (a_{c1}(h_c),\ldots,a_{cH}(h_c),z_c(h_c))\;,
\end{equation}
where $h_c$ is the audiogram value for frequency channel $c$.

The audiograms commonly contain detection thresholds for frequencies 250, 500, 1000, 2000, 3000, 4000, 6000, and 8000 Hz. These thresholds are linearly interpolated for center frequencies of the filterbank specified by $F_{cb}$. This results to values $h_1,\ldots,h_C$ which are provided as an additional input of the listener-independent variant of the dynamic processing network.

We propose to compute functions $\Theta_1(h_1),\ldots,\Theta_C(h_C)$ using a two-layer fully-connected neural network. For each frequency band $c=1,\ldots,C$ the audiogram value $h_c$ is transformed to parameters of the piece-wise linear function. These functions are represented by a fully-connected layer with $N_1$ neurons with ELU activation, followed by a fully connected layer with $H+1$ units representing the components of $\Theta_c(h_c)$.

\subsection{Convolutional NN}
\label{sec:conv}
Instead of interpretable models like the dynamic processing network presented in the previous section, a more general DNN architecture can also be used for hearing loss compensation. One possibility is to use convolutional neural networks as there have been reported to give satisfactory results for speech denoising \cite{7394335,grais2017single,park2016fully}. 

\begin{figure}
    \centering
    \includegraphics[width=7cm]{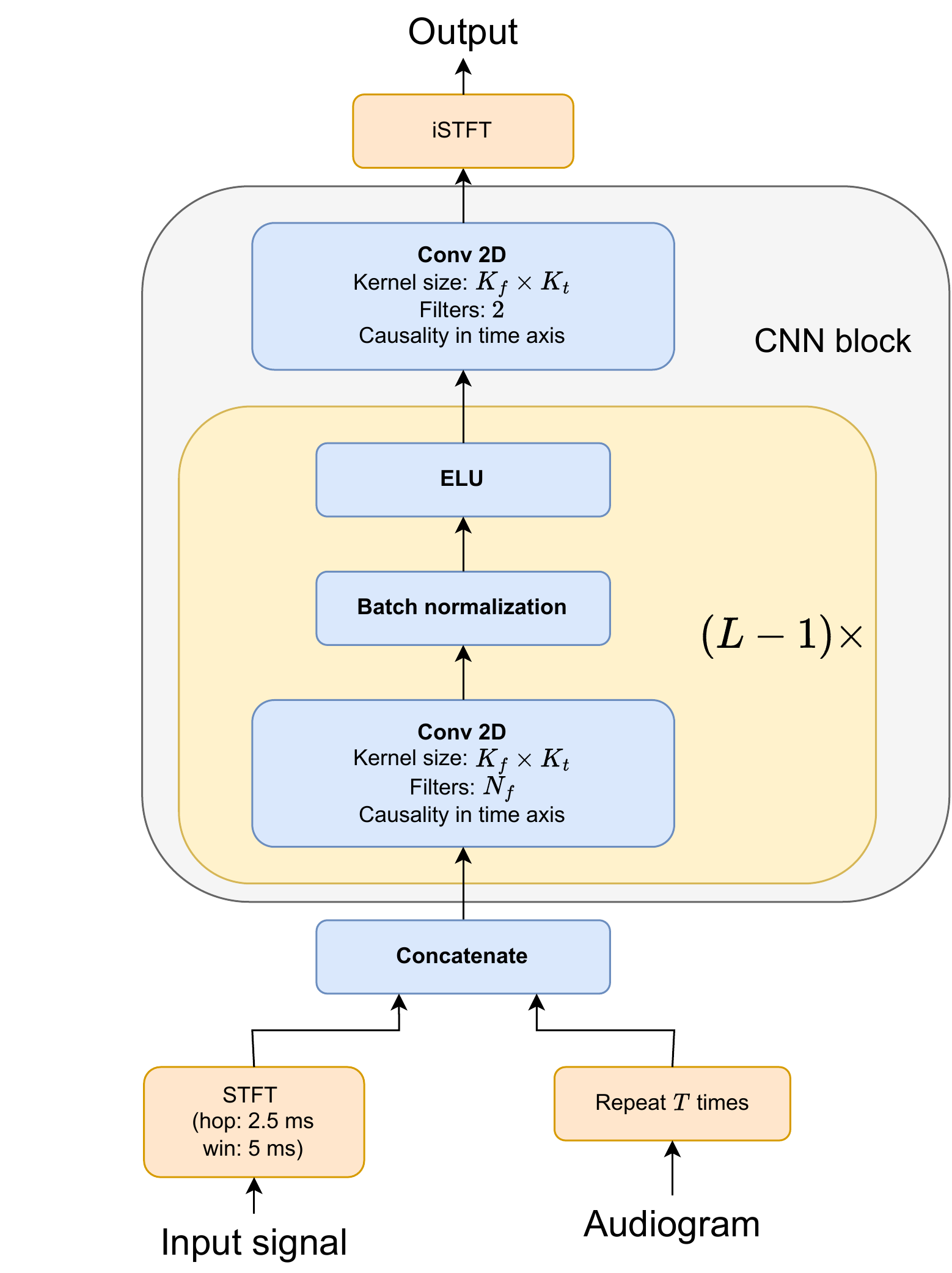}
    \caption{The structure of convolutional neural network used for hearing loss compensation}
    \label{fig:cnn}
\end{figure}

In this work, we examined a convolutional neural network which architecture is shown in Figure \ref{fig:cnn}. The inputs of the CNN-based HL compensation are an audio signal and $B$-dimensional audiogram of a listener. Next, a tensor of dimension $B\times T_{\rm comp}\times 3$ is formed that contains the complex STFT represented as a tensor of dimensions $B\times T_{\rm comp}\times 2$ concatenated with $B$-dimensional audiogram repeated $T_{\rm comp}$ times. In order to get the $B$-dimensional audiogram, an input audiogram (standard audiogram or audiogram from the Clarity Challenge\footnote{https://claritychallenge.org} dataset described in Section \ref{sec:audiograms}) with values in dB HL for certain frequencies is linearly interpolated to $B$ frequencies and its values in dB HL are transformed to the linear scale. The prepared input tensor is processed by a sequence of $L-1$ 2D convolutional layers with $N_f$ filters with kernels $K_f\times K_t$, where $K_f$ and $K_t$ are the kernel size in frequency and time dimension, respectively. After each convolutional layer, batch normalization \cite{ioffe2015batch} and exponential linear unit (ELU) nonlinearity \cite{clevert2015fast} are employed. Finally, there is the convolutional layer with two filters (for real and imaginary parts). All convolutional layers are causal in the time dimension. The output of the neural network is a tensor representing STFT of the output signal. Its dimensions are the same as of the input representation of a spectrogram: $B\times T_{\rm comp}\times 2$. Finally, iSTFT is performed to transform the HL-compensated signal to the time domain.

\section{Experiments}
\label{sec:experiments}
In this section, the HL compensation module is tested in various settings. Both the dynamic processing network and the convolutional neural network are evaluated using objective speech intelligibility metrics. 

First, the dataset and experimental setup are described. Next, the results reflecting the importance of the HL model components (spectral smearing and loudness recruitment) are presented. This is followed by tests for listener-dependent setup for standard audiograms and listener-independent setting. Finally, the results of HL compensation techniques tested separately for spectral smearing and loudness recruitment are shown. 

\subsection{Dataset}
\subsubsection{Audio}
The dataset used in the experiments comes from the Clarity Challenge \cite{graetzer2021clarity}. The speech signals that are used in the dataset come from 40 speakers. In each utterance a  talker is producing one of unique 7-10 word Clarity utterances. Only anechoic clean signals from the dataset are used to train and test the proposed method. As in this article the monaural case is investigated, only the left channel from each audio file is used. The audio signals have sampling rate of 44100 samples/s.

In order to train the hearing loss compensation models proposed in this work, the training dataset from the Clarity Challenge that contains 6000 examples (utterances of length between 1.4 and 6.4 s) was used. 
5400 files were randomly chosen from the 6000 examples to be used as a training dataset. 500 files randomly chosen from the rest were used as a validation dataset. Finally, 500 files from the Clarity Challenge development dataset were used as a test dataset.

\subsubsection{Audiograms}
\label{sec:audiograms}
\begin{figure}
    \centering
    \includegraphics[width=9cm]{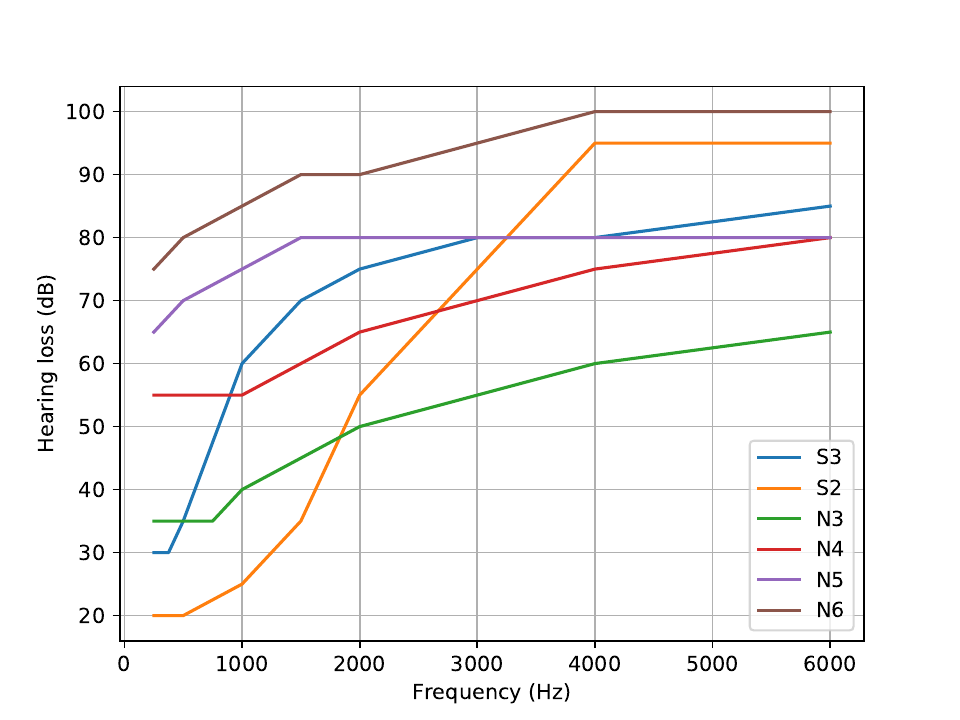}
    \caption{Standard audiograms. The values in the plots present how much is the hearing threshold above the reference value in dB HL for given frequencies (250--6,000 Hz). See also \cite{bisgaard2010standard}.}
    \label{fig:audiograms}
\end{figure}

In the case of the listener-dependent setup, standard audiograms were chosen from \cite{bisgaard2010standard}. The standard audiograms were divided into two groups according to their slope: (1) flat and moderately sloping group with identifiers N1--N6, (2) steep sloping group S1--S3. The audiograms were ordered according to hearing loss at 250 Hz, with increasing severity. The experiments were performed only for audiograms that were classified as severe according to the Clarity challenge criterion (average hearing loss between 2000 and 8000 is greater than 56 dB). Thus, we chose N3, N4, N5, N6, S2, and S3 audiograms. These audiograms are provided in Figure~\ref{fig:audiograms}. 

 We considered two main setups: listener-dependent and listener-independent.
In the listener-dependent setup the experiments are done separately for each of the standard audiograms (N3-N6, S2, S3), using 5400 training utterances, 500 validation, and 500 test utterances (as in Section V.A.1).
In the listener-independent setup, we used 90 audiograms from the Clarity Challenge that were classified as severe for training. The training data are the same 5400 utterances as in the listener-dependent case, but for each training batch, a randomly chosen audiogram from 90 severe ones is selected. The testing is done using the 500 test utterances for each standard audiogram.

\subsection{Evaluation metrics}
\label{sec:metrics}
As evaluation metrics we used STOI \cite{taal2011algorithm} and HASPI \cite{kates2014hearing}. STOI compares a signal processed by a hearing-loss compensation network and the hearing-loss model to its reference by calculating correlation between the corresponding 300 ms fragments of signals in third-octave frequency bands. The average of the correlation index is the STOI. 
We used a variant of STOI presented in Section \ref{sec:thrnoise} that takes the hearing threshold into account.

HASPI has its own hearing-loss model which incorporates peripheral hearing loss and is accurate for both normal and impaired hearing. HASPI compares the output of a hearing-loss compensation network to unprocessed reference signal. Additionally, an audiogram has to be provided at the input of HASPI as a listener characteristics, which is used in the HASPI's hearing-loss model.
The index combines measurements of envelope and temporal fine structure fidelity.

\subsection{The influence of HL model components on STOI}
In order to check the effect of the components of the hearing loss model without any compensation network, the STOI values were computed for the complete hearing loss, spectral smearing only, and loudness recruitment only conditions. In spectral smearing only, its output (${\bf x}_{\rm smeared}$) was provided at the input of the inverse outer-middle ear filter (${\bf f}_{\rm coch}$). In the case of loudness recruitment only ${\bf x}$ instead on ${\bf x}_{\rm smeared}$ was provided at the input of gammatone filterbank (Equation (\ref{eq:lrin})).

The influence of hearing loss model components was evaluated using the test data. The results of this test are summarized in Table \ref{tab:inf_of_comp}. The loudness recruitment affects the STOI in much higher extent than spectral smearing. This may be because of the clean speech material. Spectral smearing can be more susceptible in the case of noisy speech, as it makes masking more effective. 

\begin{table}[h!]
    \centering
    \caption{The influence of HL model components on STOI}
    \label{tab:inf_of_comp}
    \begin{tabular}{c|c}
    \hline
        HL model & STOI \\ \hline
        Spectral smearing only & 0.908 \\
        Loudness recruitment only & 0.531 \\
        Complete hearing loss model & 0.510 \\
    \hline
    \end{tabular}
\end{table}

\subsection{Experimental setup}
The STFT in the compensation network is calculated using a 5-ms Hann window with 2.5 ms hop size, and 512-point FFT.  

The batch size is equal to 10 and each example in the batch is a 500-frame central excerpt of the training example. These settings allow for using signals from different utterances without exceeding GPU memory.
The maximum number of epochs was set to 200, as typically after 200 epochs there was no significant improvement of STOI calculated on validation dataset. The neural network weights from the epoch that performed best on the validation set were used during the test.

The parameters of the neural network were optimized using Adam method \cite{kingma2014adam} with learning rate of 0.0001 for convolutional neural networks or 0.001 for learned dynamic processing. After each epoch, the learning rate was multiplied by 0.995. We chose the values of the learning rates, which resulted with the best STOI in validation dataset within the first 200 epochs.

In the case of listener-dependent evaluation, we computed performance metrics for the following methods:
\begin{enumerate}
 \item The case without the hearing loss compensation (No compensation). 
 \item The dynamic processing network which parameters were set according to Camfit (DPN-Camfit), tested with and without smoothing of the envelopes controlling
the gain (see Section \ref{sec:envproc}). When DPN-Camfit was tested with envelope smoothing, it was done by a 1st-order IIR filter with time constants of 10 ms for attack and 100 ms release. These settings are typical in hearing aids. 

  \item Fine-tuned dynamic processing network initialized with Camfit (DPN-FT). 
 \item Extensions of DPN-FT: with envelope processing performed by additional trainable FIR filter (DPN-FT-ENV, see Section \ref{sec:envproc}) and with optimized analysis filterbank, where $F_{cb}$ from Equations (\ref{eq:fbanal}) and (\ref{eq:fbsynth}) were fine-tuned (DPN-FT-FILT).
 \item Convolutional neural network with 6 layers (CNN).
 \item Network with encoder-decoder architecture from \cite{drakopoulos2022differentiable}, which in the original study was tested with 20 kHz sampling frequency. As the experiments in the present work has been done with 44.1 kHz sampling frequency, the direct use of the network would result with a smaller receptive field. Therefore, together with the direct use (HA-DNN), we also tested the network in which decimation is done to obtain sampling frequency which is close to the original, i.e. 22050 Hz, by adding layers with fixed weights (HA-DNN (22050)). The input to the network is decimated using a 1D convolutional layer with weights fixed to Hamming window and stride equal two. The output from the network is upsampled with transposed convolutional layer with the same weights and stride. As the center frequency of the highest band in STOI is 3810 Hz and in HASPI it is 8000 Hz, differences in sampling frequencies in the methods is not expected to affect the results. It should be also mentioned that the neural network from \cite{drakopoulos2022differentiable} uses non-causal convolutional layers and it results with a latency equal to 1904 samples. It corresponds to latency of 86 and 43 ms for sampling rate of 22050 and 44100, respectively, which is significantly higher than the latency (5 ms) of the proposed methods. 
 \end{enumerate}

In the listener-independent case we tested the following: 
\begin{enumerate}
\item The dynamic processing network with parameters obtained by transforming an audiogram using fully-connected layers (DPN-LI) (see \ref{sec:audiogramdepend}). This variant does not use Camfit procedure,
 \item The convolutional neural network with 4, 5, and 6 levels (CNN-LI).
 \item Dynamic processing network followed by the convolutional neural network, where the representation $Y$ (see Equation \ref{eq:outputstft}) is provided at the input of the CNN presented in Section \ref{sec:conv} (DPNCNN-LI). 
 \item Network with encoder-decoder architecture from \cite{drakopoulos2022differentiable} (HA-DNN-LI).
\end{enumerate}
 When the hearing loss compensation was done with the dynamic processing network the layers that transform an audiogram into parameters $\Theta$ had $N_1=80$ neurons. The kernel size in the convolutional layers is of $7\times 3$ (the average STOIs on the validation dataset in the experiment described in Section \ref{sec:ks} did not change significantly with the size of kernel -- 0.909, 0.912, 0.915, for $7\times 3$, $7\times 5$, and $7\times 20$, respectively. Therefore, we chose the kernel with the smallest number of computational operations). Each hidden convolutional layer ($l=1,\ldots,L-1$) has 48 filters.

\subsection{Listener-dependent results for standard audiograms}
\label{sec:ldresults}
\begin{figure*}[h]
    \centering
    \includegraphics[width=18cm]{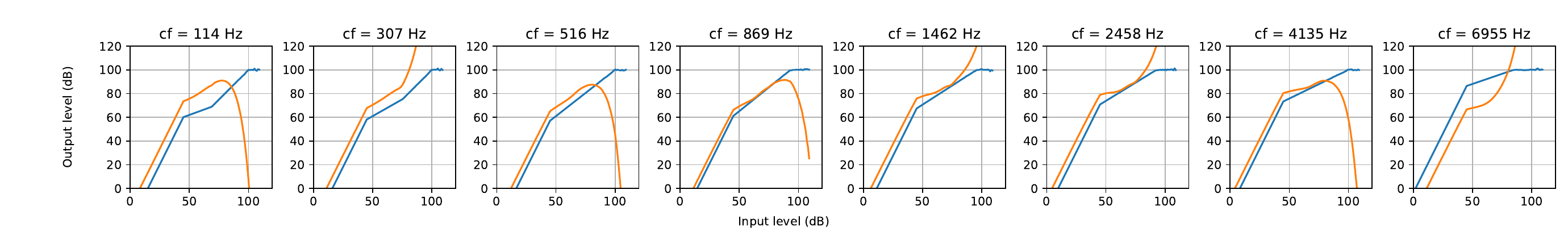}
    \caption{The input-output characteristics of the dynamic processing network. Parameters of the curves are frequencies. The blue lines represent input-output characteristics for compressor obtained with Camfit procedure. The orange lines are after fine-tuning.}
    \label{fig:inout}
\end{figure*}

The results of the dynamic processing neural network described in Section \ref{sec:dp} are presented in Table \ref{tab:ld}. The STOIs and HASPIs presented in the table are the averages over all used standard audiograms (N3, N4, N5, N6, S2, and S3). 

The DPN-Camfit gives a significant benefit in comparison to the case without the hearing loss compensation. The addition standard envelope smoothing used in hearing aids decreases STOI and HASPI. The fine-tuning (DPN-FT) gives improvement in terms of STOI and HASPI in comparison to Camfit. The performance achieved after fine-tuning of the basic variant to the dynamic processing network can be further improved by fine-tuning the filterbank $F_{cb}$ in DPN-FT-FILT. In the case of variant with training of the envelope processing filters (see Section \ref{sec:envproc}) DPN-FT-ENV brought slight improvement in comparison to the basic variant only for less severe audiograms (N3, N4). The CNN outperformed DPN-Camfit but not its fine-tuned versions. HA-DNN gave similar results to CNN, slightly better in terms of STOI, but worse in terms of HASPI.

In order to check the statistical significance of the results with HASPI values, the Wilcoxon signed-rank with continuity correction was performed, using its implementation in R package. The improvement of HASPI values of DPN-Camfit fine-tuned to get the DPN-FT turned out to be statistically significant (p-value $<$ 0.001). The improvement of HASPI for DPN-FT over HA-DNN was also statistically significant with p-values smaller than 0.001. However, this was not a case for CNN.

\begin{table}[h]
    \centering
    \caption{Average STOIs and HASPIs of evaluated listener-dependent methods for standard audiograms. The last column shows whether the improvement of the DPN-FT over other systems is statistically significant. Triple asterisks mean $p<0.001$}
    \label{tab:ld}
    \begin{tabular}{l|r|rr}
    \hline
        Method & STOI & HASPI & signif. \\ \hline
        No compensation & 0.510 & 0.368 & ***\\
        DPN-Camfit & 0.803 & 0.947 & ***\\
        DPN-Camfit (envelope proc. 10/100 ms) & 0.766 & 0.931 & ***\\
        DPN-FT & 0.887 & {\bf 0.966} & \\
        DPN-FT-ENV & 0.864 & 0.920 & ***\\
        DPN-FT-FILT & {\bf 0.889} & 0.960 & ***\\
        CNN & 0.843 & 0.833 & \\ 
        HA-DNN & 0.772 & 0.660 & ***\\
        HA-DNN (22050) & 0.858 & 0.821 & ***\\\hline
    \end{tabular}    
\end{table}

The influence of the size of kernel of filters in convolutional layer that process envelopes in DPN-FT-ENV is summarized in Table \ref{tab:env_ker_size}. The difference between 20 and 50 taps is small in the case of less severe audiograms (N3, S2), but for more severe hearing losses, bigger kernel size results to worse STOI.
\begin{table}[h!]
    \centering
    \caption{STOIs of DPN-FT-ENV, with different sizes of envelope filter}
    \label{tab:env_ker_size}
    \begin{tabular}{c|cc}
    \hline
        Audiogram & 20 taps  & 50 taps \\ \hline
        N3 & 0.927 & 0.923 \\
        N4 & 0.912 & 0.892 \\
        N5 & 0.860 & 0.797 \\
        N6 & 0.696 & 0.479 \\
        S2 & 0.896 & 0.906 \\
        S3 & 0.895 & 0.863 \\ \hline
    \end{tabular}
\end{table}

The comparison of input-output characteristics of the learned dynamic processor to Camfit characteristics are shown in Figure \ref{fig:inout}. The last band  with frequencies above 7 kHz is not shown, as the DPN-FI for these frequencies does not transmit signal. For higher levels (the levels that did not occur in the training dataset) the signal is significantly attenuated. For input levels between 50 and 80 dB there are greater output levels for frequencies below 1 kHz, while for higher frequencies they are similar or smaller.

\subsection{Listener-independent results}
The results of listener-independent methods are presented in Table \ref{tab:li-stoihaspi}. 
\begin{table}[h]
    \centering
    \caption{Average STOIs and HASPIs of evaluated listener-independent methods. The last column shows whether the improvement of the DPNCNN-LI over other systems is statistically significant. Triple asterisks mean $p<0.001$}
    \label{tab:li-stoihaspi}
    \begin{tabular}{c|c|cc}
    \hline
         & STOI & HASPI & signif. \\ \hline
        DPN-LI & {\bf 0.829} & 0.933 & *** \\
        CNN-LI (4 layers) & 0.731 & 0.856 & ***\\
        CNN-LI (5 layers) & 0.764 & 0.915 & ***\\
        CNN-LI (6 layers) & 0.767 & 0.918 & ***\\ 
        DPNCNN-LI  & 0.823 & {\bf 0.957} & \\ 
        HA-DNN-LI & 0.645 & 0.666 & ***\\
        HA-DNN-LI (22050) & 0.647 & 0.612 & ***\\ \hline
    \end{tabular}
\end{table}

Based on STOI, DPN-LI outperforms DPN-Camfit and all tested convolutional neural networks, as well as the combination of the dynamic processing network and convolutional layers (DPNCNN-LI). CNN-LI performed better than HA-DNN-LI. 

The HASPI values, in general, are consistent with STOI metric (see Table \ref{tab:li-stoihaspi}), though in terms of HASPI, the best performing system is DPNCNN-LI. We also tested the HA-DNN with weights made available by its authors\footnote{https://github.com/HearingTechnology/DNN-HA} \cite{10141861}. This network was trained using TIMIT corpus and operates at 20 kHz sampling frequency, and therefore we downsampled our test dataset to 20 kHz. This network resulted in HASPI 0.420, which is lower than the results in Table IV, which can be explained by mismatch in the training data. 

The statistical significance of the results for listener-independent methods were tested using the Wilcoxon signed rank test with continuity correction, similarly as in Section \ref{sec:ldresults}. The test showed that the improvement of the best-performing listener-independent system (DPNCNN-LI) over all other systems was statistically significant in all cases (p-value $<$ 0.001).  

 The performance of the listener-independent methods in Table \ref{tab:li-stoihaspi} can be compared to the listener-dependent methods in Table \ref{tab:ld}. DPN-LI and DPNCNN-LI performed better than DPN-Camfit in terms of STOI. However, its performance is worse in comparison to the learned listener-dependent methods. 

\subsection{Efficiency of compensation for individual components of the hearing loss model}
\label{sec:ks}
In order to evaluate the efficiency of spectral-smearing compensation, the loudness recruitment module in the hearing loss model was turned off. The convolutional neural network described in Section \ref{sec:conv} was tested for different kernel sizes: $7\times 3$, $20\times 3$, and $7\times 5$, and different number of layers (3, 4, 5, 6). The number of filters in each layer from 1 to $L-1$ was 48.

The results for spectral smearing are shown in Table \ref{tab:ss}. Generally, it is apparent that more complex networks (more layers and parameters) give better performance in terms of STOI. The best results were obtained for 6 layers with $20\times 3$ kernel, with STOI of $0.931$ what clearly outperforms the case without any hearing loss compensation with STOI of $0.908$. From the Table \ref{tab:ss} it can be noticed that the kernel size has smaller influence than the number of layers.

\begin{table}[h!]
    \centering
    \caption{STOIs of the CNN-based compensation of spectral smearing}
    \label{tab:ss}
    \begin{tabular}{r|rrr}
\hline
          layers &    7x3 &   20x3 &    7x5 \\
\hline
               3 &  0.909 &  0.914 &  0.906 \\
               4 &  0.908 &  0.920 &  0.916 \\
               5 &  0.923 &  0.924 &  0.913 \\
               6 &  0.922 &  {\bf 0.931} &  0.917 \\ \hline
 no compensation &  0.908 &     &     \\
\hline
\end{tabular}
\end{table}

In the case of the HL model, in which spectral smearing is turned off, both learned dynamic processing and convolutional neural networks has been tested. The results are presented in Table \ref{tab:lr}. Increasing the number of convolutional layers leads to STOI improvement. The learned dynamic processing gives better STOI than the convolutional neural network with 6 layers.
\begin{table}[h!]
    \centering
        \caption{STOIs of the CNN-based and DPN-LI compensation of loudness recruitment}
    \label{tab:lr}

    \begin{tabular}{l|r}
\hline
                     layers &   STOI \\
\hline
            no compensation &  0.531 \\
            CNN-LI 4 layers &  0.738 \\
            CNN-LI 5 layers &  0.762 \\
            CNN-LI 6 layers &  0.792 \\
 DPN-LI &  0.863 \\
\hline
\end{tabular}
\end{table}

\section{Conclusions}
\label{sec:conclusions}
In this paper the problem of designing the hearing-loss compensation neural network has been addressed. 
 Both listener-dependent and listener-independent results show that fine-tuning of the dynamic processing network with weights initialized according to Camfit procedure leads to improvement of the predicted speech intelligibility in comparison to fixed Camfit settings.
The best results for listener-independent setup were obtained using combination of dynamic processing network and CNN.
 The experiments in which compensation techniques were applied to the components of the hearing loss model, show that the convolutional neural network can partially compensate spectral smearing. It was also found that both convolutional neural network and learned- dynamic processing network can partially compensate loudness recruitment. The computationally efficient learned dynamic processing outperforms CNN with 6 convolutional layers.

\section*{Appendix}
\subsection{Outer-middle ear filter}
The middle-outer ear frequency characteristics is taken from \cite{shaw1974transformation}. In order to simulate filtering by outer and middle ear, FIR linear phase filter was designed from the middle-outer ear frequency characteristics using window method with Kaiser with parameter 4.  In the case of its inverse, which is applied as one of the final step in the hearing loss model, the middle-outer ear frequency characteristics in dB was multiplied by -1. Next, the filter was designed as in the case of filtering from outer to inner ear.

\subsection{Smearing matrix}
The smearing matrix is used to transform frames of the magnitude spectrogram of the input signal according to Equation (\ref{eq:smearing}). After the transformation the spectrum is smeared like it would be processed by broader auditory filters. The smearing matrix can be computed as
\begin{equation}
    {\bf S}={\bf S}_{\rm normal}^{-1}{\bf S}_{\rm wide}
\end{equation}
where the rows of ${\bf S}_{\rm normal}$ contain magnitude responses of gammatone filters with normal widths, ${\bf S_{\rm wide}}$ contains widened filters.  
The gammatone magnitude response can be written as
\begin{equation}
    W(g)=(1+pg)\exp(-pg)\;,
\end{equation}
where $g$ is deviation from the center frequency of the filter divided by the center frequency, while $p$ is a pass-band parameter $(q{\rm ERB})/f=4/p$, and $q$ is a widening factor in comparison to the normal filter \cite{moore1983suggested}. 

\subsection{Gammatone filterbank}
The gammatone filterbank is used in the loudness recruitment component of the hearing loss model as described in Section \ref{sec:lr}.
The filters  were chosen to have widths and shapes comparable to those of auditory filters measured in subjects with severe cochlear hearing loss \cite{glasberg1986auditory}.

Additionally, after each bandpass gammatone filter, an infinite impulse response high-pass filter was applied to reduce the effect of the ‘‘tail’’ of the low-frequency side of the filters with the four highest center frequencies.

\subsection{Temporal envelope extraction filters}
The temporal envelope extraction filters are used in the loudness recruitment component of the hearing loss model described in Section \ref{sec:lr}. In order to extract envelopes in frequency bands, 2nd order low-pass ellipitic filters were used. Their cutoff frequencies denpend on center frequency of a band.

Ripple parameters of the elliptic filters were set to  $0.25$ for the pass band and $35$ stop band. These filters applied to in both directions to Kronecker delta to get 1660 samples of impulse response. Next, this impulse response was downsampled to match samplling rate of the processed envelopes.





\ifCLASSOPTIONcaptionsoff
  \newpage
\fi



\bibliographystyle{IEEEtran}
\bibliography{bibliography}
\end{document}